\documentstyle[preprint,aps]{revtex}
\begin{document}
\draft

\title{Phase Locking, Devil's Staircases, Farey Trees, and Arnold Tongues
 in Driven Vortex Lattices with Periodic Pinning}
\author{C.~Reichhardt and Franco Nori}
\address{Department of Physics, The University of Michigan,
Ann Arbor, Michigan 48109-1120}

\date{\today}
\maketitle
\begin{abstract}
Using numerical simulations, we observe 
phase locking, Arnold tongues, and Devil's staircases 
for vortex lattices driven at varying angles with respect 
to an underlying superconducting periodic pinning array.  
This rich structure should be observable in transport measurements.
The transverse $V(I)$ curves have a Devil's staircase structure,
with plateaus occurring near the driving angles 
along symmetry directions of the pinning array.
Each of the plateaus corresponds to a different dynamical phase 
with a distinctive vortex structure and flow pattern.
\end{abstract}
\vspace*{-0.1in}
%
\vspace*{-0.4in}
%
\narrowtext

{\it Introduction.---} Numerous nonlinear driven systems in 
physics, astronomy, and engineering exhibit striking responses with 
complex phase-locking plateaus characterized by Devil's staircases, 
Arnold tongues, and Farey trees\cite{devil,notation1,notation2,tongues}.  
Here, we present the first evidence that these structures 
can be observed in bulk superconductors.

Driven vortex lattices (VLs) interacting with either random or 
periodic disorder have attracted growing interest due to the
rich variety of nonequilibrium dynamic phases which are 
observed in these systems.
These phases include the elastic and plastic 
flow of vortices which can be related
to VL order and transport properties
\cite{dynamic,barrier,exp,DrivenShort,paps}.
  Periodic pinning arrays interacting with VLs
are now attracting increasing attention as recent
experiments with patterns of holes \cite{holes} and
magnetic dots \cite{Schuller}
have produced interesting commensurability effects 
and enhanced pinning.
  These systems are an excellent realization of an
elastic lattice interacting with a periodic substrate 
that is found in a wide variety of condensed matter systems 
including charge-density-waves, Josephson-junction arrays, 
and Frenkel-Kontorova-type models of friction (see, e.g., \cite{Gorkov}).
An interesting aspect of periodic pinning arrays that has not 
been addressed so far is how the symmetry properties of the
array affect the transport properties as the VL is driven at
different angles.

We find that as a slowly increasing transverse force is 
applied to a VL already moving in the longitudinal direction, 
the VL undergoes a remarkable series of {\it locking transitions\/} that 
significantly affect both the VL ordering and transport properties.
These locking phases occur when the direction of the vortex motion 
locks with a symmetry direction of the pinning array. As the VL
passes through these phases, the transverse velocity component as 
a function of increasing transverse drive shows a series of
plateaus which form a Devil's staircase structure 
\cite{devil,notation1,notation2}.
At the boundaries of certain locked phases the VL
undergoes a transition to a
{\it plastic flow} phase in which defects are generated in the VL.
In the locked phases the VL undergoes
{\it elastic} flow in static 1D channels and the overall VL
has a variety of orderings, including triangular and square.

{\it Simulation.---} We consider a 2D slice of $N_{v}$ 
3D rigid vortices 
interacting with 
a square array of $N_{p}$ parabolic wells, 
with lattice constant $a$, and periodic boundary conditions.
We integrate \cite{DrivenShort} the 
equations of vortex motion
$ {\bf f}_{i} = {\bf f}_{i}^{vv} + {\bf f}_{i}^{vp} + {\bf f}_{d}
= \eta{\bf v}_{i}$.
The total force $ {\bf f}_{i}$
on vortex $i$ includes interactions with
other vortices ${\bf f}_{i}^{vv}$, pinning
${\bf f}_{i}^{vp}$ by parabolic wells, 
and an applied driving force 
${\bf f}_{d} = f_x {\bf \hat{x}}$.
The vortex-vortex interaction between vortex $i$ and 
the other $N_v$ vortices is
${\bf f}_{i}^{vv} = \sum_{j=1}^{N_{v}}f_{0}K_{1}(|{\bf r}_{i} - {\bf r}_{j}|/
\lambda){\bf {\hat r}}_{ij}$, where $K_{1}(r/\lambda)$ is a modified
Bessel function, $\lambda$ is the penetration depth,
$f_{0} = \Phi_{0}^{2}/8\pi^{2}\lambda^{3}$, ${\bf {\hat r}}_{ij} =
({\bf r}_{i} - {\bf r}_{j})/|{\bf r}_{i} - {\bf r}_{j}|$, and
we set $\eta=1$.
All lengths, fields, and forces are given in units of $\lambda$,
$\Phi_{0}/\lambda^{2}$, and $f_{0}$, respectively. For most of the
results presented here
the number of vortices is close to the number of pinning sites,
$N_{v} = 1.062 N_{p}$.
We have conducted a series of simulations with different 
pinning parameters so that accurate phase diagrams 
of the dynamic phases can be obtained.
In order to investigate finite size effects we
have examined system sizes varying between
$36\lambda\times 36\lambda$ and
$108\lambda\times108\lambda$, with
$N_{v}$ between
$N_{v} = 550$ and $N_{v} = 4955$.

{\it Voltage--Current Response.---} First, 
the VL ground state at zero applied driving force is 
found by simulated annealing 
(i.e., by cooling the VL from high-$T$).
After a low energy ground state is found, a
slowly increasing driving force, 
$f_{x}$, is applied along the horizontal
symmetry axis of the square pinning.
We find that increasing $f_{x}$ in
increments of $0.001f_{0}$ every 400 MD steps,
from $f_{x} = 0$ to $f_{x} = 3.0f_{0}$, 
is slow enough that the vortex dynamics does not depend
on the rate of increase of $f_{x}$.
Once $f_{x}$ is brought to
$3.0f_{0}$ it is held constant while
a force, which we label $f_{y}$, is applied in the transverse
or $y$-direction.
We increase $f_{y}$ from $0$ to $3.25f_{0}$, also in 
increments of $0.001f_{0}$ every 400 MD steps.
The total driving force has a net magnitude of
$f_{d} = (f_{x}^{2} + f_{y}^{2})^{1/2}$ at an angle
$ \theta = \tan^{-1}(f_{y}/f_{x})$ 
with respect to the $x$-direction.
We compute the average velocity of the 
moving vortices in both the longitudinal
$V_{x} = (1/N_{v})\sum_{i=1}^{N_{v}}{\bf v}_{i}\cdot{\bf {\hat x}}$
and the transverse
$V_{y} = (1/N_{v})\sum_{i=1}^{N_{v}}{\bf v}_{i}\cdot{\bf {\hat y}}$ 
direction, as $f_{y}$ is increased. 
Velocity versus driving plots 
correspond to experimentally measurable voltage-current $V(I)$ curves.

\vspace*{-0.1in}

In Fig.~1(a) we present a typical plot of
$V_{x}$ and $V_{y}$.
For $f_{y} \lesssim 0.4f_{0}$, $V_{y} = 0$ indicating that
the VL is {\it pinned} in the $y$-direction
even though the VL is moving in the $x$-direction.
Depinning in the transverse direction
occurs at $f_{y} = 0.4f_{0}$, as indicated
by the sharp jump up in $V_{y}$.
We label this
critical transverse depinning force $f_{y}^{c}$.
A jump up in $V_{x}$ is also observed at $f_{y}^{c}$.
As $f_{y}$ is linearly increased,
$V_{y}$ does {\it not\/} grow linearly
but instead in a remarkable series of {\it jumps\/} and 
{\it plateaus\/} of varying sizes \cite{notation2}.
Along the plateaus $V_{y}$ is constant or increasing very
slowly, indicating that the vortex motion is {\it locked\/} 
in a certain direction for a finite range of increasing $f_{y}$.
The small jumps and dips in $V_{x}$ correspond
to the onset of plateaus in $V_{y}$.
The plateaus in $V_{y}$ occur
when the
ratio of $f_{y}$ to $f_{x}$ is near a rational value:
$f_{y}/f_{x} = p/q$,
where $p$ and $q$ are integers.
In Fig.~1(a) the largest plateaus occur
at $p/q = 0,1/3, 1/2, 2/3$ and $1$.
Fig.~1(b) shows a blow-up of a region in Fig.~1(a) for values of
$f_{y} = 0.6f_{0}$ to $ 2.1f_{0}$, where additional plateaus
at $p/q = 1/5, 1/4, 2/5, 3/7$, and $3/5$ are highlighted.
For larger system sizes we find
exactly the same behavior in $V_{y}$ and
$V_{x}$ as observed in Fig.~1,
indicating that 
it is independent of the system size.

{\it Vortex Dynamics and the Origin of the Plateaus.---}
To understand why the plateaus occur as well as the VL 
dynamics in the plateau and non-plateau regions,
in Fig.~2(a-d) we plot the vortex trajectories for
rational ratios of $f_{y}/f_{x} = 0, 1/2$, $1$, and
the irrational ratio $f_{y}/f_{x} = 2\pi/11 = 0.571... $
In Fig.~2(a), where 
$f_{y} < f_{y}^{c}$, the vortex motion traverses pin sites
periodically and it is only along the $x$-direction---with 
the vortex flow restricted in
1D paths {\it along\/} the pinning rows.  This periodic 1D 
motion persists up to $f_{y} = f_{y}^{c}$, at which point the 
vortices also begin to flow in the $y$-direction.  
In Fig.~2(b), for $p/q = 1/2$ where a large 
plateau in $V_{y}$ is observed in Fig.~1, the 
vortices again exhibit periodic motion and flow in 
1D channels {\it along} the pinning sites---and along a 
symmetry axis of the pinning array at an angle 
$\theta = \tan^{-1}(1/2)$ from the $x$-axis. 
A similar periodic 1D motion is seen in (d) for $f_{y}/f_{x} = 1$, 
with the VL motion at $45^{\rm o}$ from the $x$-axis.
In Fig.~2(c), at the irrational $f_{y}/f_{x}$ ratio, 
the vortex trajectories are 
different than those observed in Fig.~2(a,b,d).
Here the quasiperiodic vortex trajectories drift over time,
eventually covering the sample (i.e., ergodic-like motion).
In general, the plateau regions (with rational $f_{y}/f_{x}$) 
in $V_y(I)$ correspond to periodic 1D vortex trajectories, 
while the non-plateau regions produce quasiperiodic 
trajectories\cite{devil,notation1,Nori}.

To understand how the vortex motion locks into certain 
driving angles, we first consider the case
$f_{y}/f_{x} = 0$. Here the vortices move along the pinning
rows in 1D paths, with each vortex traversing a distance 
$a - 2r_{p}$ between pinning sites, as seen in Fig.~2(a).
An application of a transverse force
$f_{y}$ causes the moving vortices to drift a
small distance in the $y$--direction.
Once the vortices interact with the pinning sites, 
they feel a force that moves them towards the center 
of the pinning site which keeps them locked
along the $x$-direction.  When $f_{y}$ is large enough, 
$f_y^c \gtrsim f_x \, \tan (r_p/a) \; $ \cite{Nori}, 
the vortices are able to break off from moving only along the
$x$-direction and start moving in the $y$-direction as well.

As $f_{y}$ is increased beyond $f_{y}^{c}$, the net
driving force vector will be at an angle with the horizontal.
Due to the symmetry of the square pinning array, along the angles where
$\theta = \tan^{-1}(p/q)$, the vortices encounter pinning sites
{\it periodically\/} spaced a distance $a_{\theta}$ apart.
This distance is related to the pinning
lattice constant $a$ by $a_{\theta} = a(p^{2} + q^{2})^{1/2}$.
Along these commensurate angles, the vortex motion will 
be periodic and locked in 1D channels 
in a similar manner as the $f_{y}/f_{x} = 0$ case.
The force needed to depin the vortices from the commensurate 
angles will vary since $a_{\theta}$ varies. 
For values where $a_{\theta}$ is small, the vortices will
move only a small distance between pinning sites, so
a higher depining force is needed.
For large $a_{\theta}$ the vortices will move a much 
longer distance before encountering the pinning sites, 
so a much smaller depinning force is needed.
This is in agreement with Fig.~1 where the {\it largest\/} plateaus
(due to enhanced pinning) occur for values of $p/q$ that
produce the lowest distance between pinning sites, that is the 
{\it smallest\/} $a_{\theta}$ 
(i.e., $p/q = 0/1$, $1/1$, and $1/2$).

The onset of certain plateaus coincide with a variety of 
{\it structural} transitions in the VL. We quantify this
angle-dependent evolving {\it topological order\/}
by using the Voronoi (or Wigner-Seitz) construction to obtain 
the fraction of vortices with coordination numbers
six, $P_{6}$, and four, $P_{4}$. 
In Fig.~3(a,b) we show the evolution of $P_{6}$ and $P_{4}$ as $f_{y}$
is increased, for the same system as in Fig.~1.
For $f_{y} < f_{y}^{c}$, $P_{6} \approx 0.68$, indicating a
mostly triangular VL.
At $f_{y} = f_{y}^{c}$, a dip in $P_{6}$, along with
direct observation of the VL flow,
show that the VL {\it disorders}
due to {\it plastic\/} deformations.
Right after the initial dip in $P_{6}$ the VL suddenly regains 
considerable triangular ordering, as indicated by 
$P_{6} \approx 0.95$.  Small dips in $P_{6}$ can be seen near 
the $1/4$, $1/3$, and $2/3$ locking regions.
At the $1/2$ locking region the VL is considerably disordered, as indicated
by the sharp drop in $P_{6}$.  This is consistent with Fig.~3(c), 
where both the vortex positions and Voronoi polygons are shown
for a $12\lambda\times12\lambda$ region
in the $1/2$ locking region. At the
$1/1$ locking region $P_{6}$ drops almost to zero while
$P_{4}$ increases to about $0.9$, indicating a structural phase
transition from a triangular to a {\it square} VL.
Here, the $f_y = f_x$ symmetric drive is what 
produces a moving square VL.
The less symmetric drives 
($2 f_y = f_x \,$ and $3 f_y = f_x$),
produce more distorted squares \cite{Nori}.
For the special case when $f_y = 0$ and for the $B$ used in Fig.~3, 
correlations between nearby VL rows are strong, 
and near 2/3 of the VL has triangular order
(which diminishes for weaker $B$'s). 
In Fig.~3(d,e) the vortex positions and Voronoi polygons 
are shown for (d) right {\it before\/} the transition to the 
$1/1$ locking region and (e) {\it in\/} the $1/1$ locking region 
showing the triangular and square ordering of the VL respectively.
Right {\it at\/} the boundaries of the $1/1$ phase, the VL is strongly 
disordered and has a similar structure to Fig.~3(c).

%

\vspace*{-0.1in}

{\it Phase Diagrams with Arnold Tongues.---} We have derived 
{\it five phase diagrams\/} which indicate the evolution of 
the plateau regions versus the following parameters: 
$f_{p}$, $n_p$, $r_p$, commensurability, and disorder.
These five phase diagrams are all very similar, and thus here 
we present only one: Fig.~4(a).  This is obtained by 
conducting a series of simulations in which the 
maximum pinning force $f_{p}$ is varied
between $0.25 \leq  f_{p}/f_{0} \leq 2.75$.  The
phase diagram shows 18 clearly defined (shaded)
{\it Arnold tongues\/} or plateaus 
\cite{devil,notation1,notation2,tongues}. 
As $f_{p}$ is decreased the widths of the tongues
also show a corresponding decrease.
For $f_{p}/f_{0} > 2.5$ several locking phases are lost
(i.e., 1/6, 4/7, 5/6) due to overlapping by other 
locking regions. For $f_{p}/f_{0} < 1.0$
only the strongest plateau regions can be resolved
within the accuracy of our calculations.

The phase diagram in Fig.~4(a) has the {\it same\/} structure 
as Arnold tongues \cite{devil,notation1,notation2,tongues} 
found in phase locking systems where the widths of 
the tongues, or locking regions, increase as the 
nonlinear coupling increases.
Here, the coupling is between the vortices and 
the pinning array, and is increased with increasing 
$f_{p}$, $r_{p}$, $n_p$, density of vortices 
(i.e., the commensurability $B/B_{\phi}$), 
and pin-location order \cite{Nori}.
In Fig.~4(b) we present the width of the $0/1$ locking region
for varying pinning density in units of the pinning lattice 
constant $a$. 
As $a$ decreases the width of the locked region increases.
This can be understood by considering that as
$a$ decreases the vortices in the locked region will
move a smaller distance between pinning sites; 
thus a higher transverse force is needed to break
the vortices away from the locked region. The widths of the
other locked regions show the same behavior as the $0/1$ region
for increasing $a$ \cite{Nori}.

We have also examined the effects of pin disorder
on the width of the locking regions by conducting a series of
simulations in which the pinning sites are randomly
displaced up to an amount $\delta r$ away from the
perfectly square pinning lattice.
We consider the case where $\delta r = a/2$ to be a
good approximation to a random pinning array.
In Fig.~4(c), 
we examine how the width of the $0/1$ locking region, 
$f^c_y$, decreases as $\delta r$ is increased.
It is of interest to compare our results for large 
disorder with Ref \cite{barrier}(a) in which a nonzero 
transverse critical force $f^c_y$ was predicted.
%
%
Recent $T= 0$ MD simulations have observed extremely 
small transverse barriers \cite{barrier}(b).
We find that for large disorder, $\delta r = a/2$,
a true transverse barrier (i.e., $V_{y} = 0$)
is {\it not\/} observed.
Also, for a triangular array of pins, 
the plateaus occur for 
$\theta = \tan^{-1}(\sqrt3p/(2q + 1))$.


{\it Summary.---} In conclusion, we have found that as an 
increasing transverse force is applied to a strongly driven 
VL interacting with a periodic pinning array, 
the VL undergoes a remarkable series of locking
transitions in which both the VL order and
flow patterns change.
As the VL passes through these
transitions, $V_{y}$ exhibits
a striking series of plateaus forming a Devil's
staircase structure.
The width variations of these plateaus
with different pinning form Arnold tongues which can be
indexed via a Farey tree construction.
These locking effects occur whenever the VL 
is driven along a symmetry angle of the pinning array.
For a square pinning array, the locking phases occur when
driving in the longitudinal direction is
a rational ratio, $f_{y}/f_{x} = p/q$.
These predictions can be tested experimentally and we hope
that this work will motivate several novel experiments.
Moreover, other candidate systems where these predictions 
may be accessible include: driven Wigner crystals 
interacting with a periodic array of donors, 
driven colloids interacting with optical-trap arrays,
spin- and charge-density waves,
Josephson-junction arrays, and solid friction experiments.

We thank F.~Marchesoni, M.~Bretz,
and very specially C.J.~Olson for useful discussions.

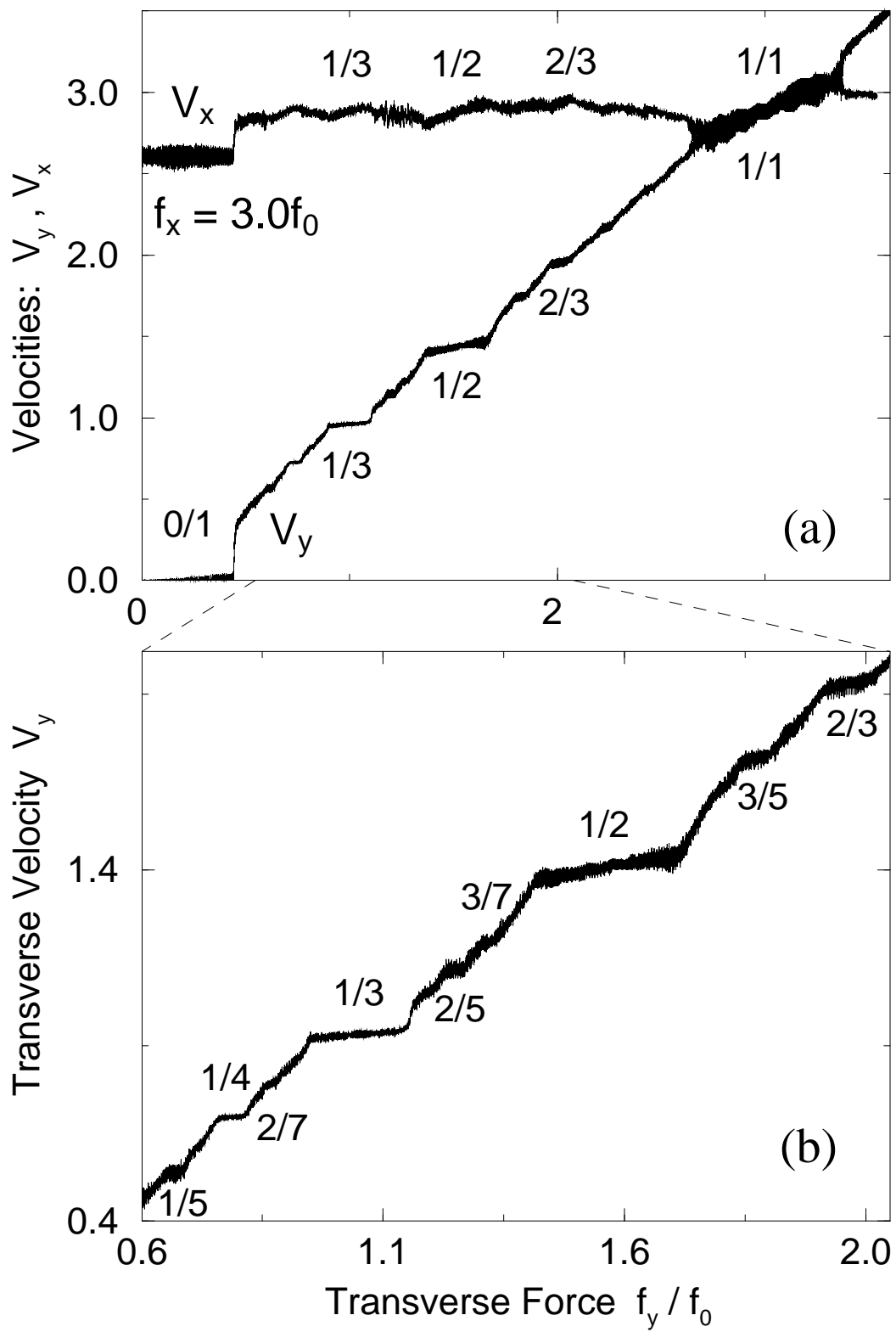
\begin{figure}
\vspace*{1.3in}
\caption{
(a) Average longitudinal $V_{x}$ (upper curve) and 
transverse $V_{y}$ velocities versus the transverse
driving force $f_{y}$ for a $36 \lambda \times 36 \lambda $
sample with a square pinning array, desity of field lines $B$
satisfying $B/B_{\phi} = 1.062$, matching field 
$B_{\phi} = 0.4 \, \Phi_{0}/\lambda^{2}$, 
density of pinning sites $n_{p} = 0.4 /\lambda^{2}$,
$f_{p} = 2.5f_{0}$, 
$a=1.57\lambda$, and $r_{p} = 0.3\lambda$. 
$f_x$ is fixed at $f_{x} = 3.0f_{0}$. Plateaus are seen in
$V_{y}$ near values where $f_{y}/f_{x} = p/q$, where $p$ and $q$ are
integers. The largest plateaus (at $0/1$, $1/3$, $1/2$, $2/3$, and $1/1$) 
 are clearly seen. 
(b) shows a blow up of $V_{y}$ from (a), for $f_{y} = 0.6f_{0}$ to
$2.1f_{0}$, where additional plateaus at $1/5$, $1/4$, $2/5$, $3/7$, and
$3/5$ can be seen more clearly. The overall structure in $V_y$ 
is that of a Devil's staircase [1-3].
}
\label{fig1}
\end{figure}

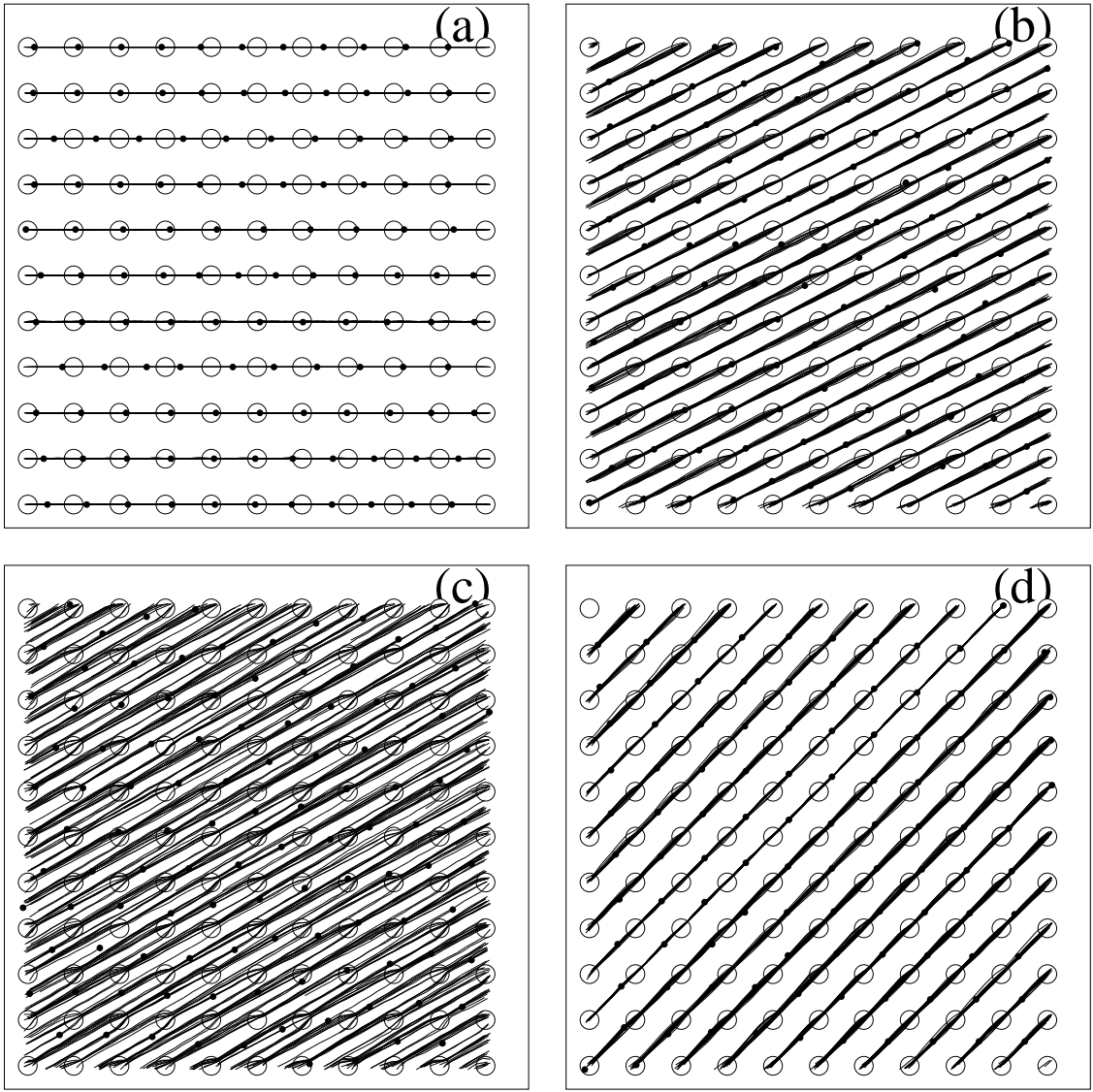
\begin{figure}
\vspace*{1.0in}
\caption{
The vortex trajectories for a subset of 
the system in Fig.~1 at the plateau regions
(a) $f_{y}/f_{x} = 0$, (b) $f_{y}/f_{x} = 1/2$, and
(d) $f_{y}/f_{x} = 1/1$, and at the non-plateau region (c)
$f_{y}/f_{x} = 2\pi /11 = 0.571... \ $  
At the plateau regions, 
the vortices move in 1D channels, periodically along the 
pinning rows; while at the non-plateau regions the
vortices exhibit quasiperiodic trajectories.
}
\label{fig2}
\end{figure}

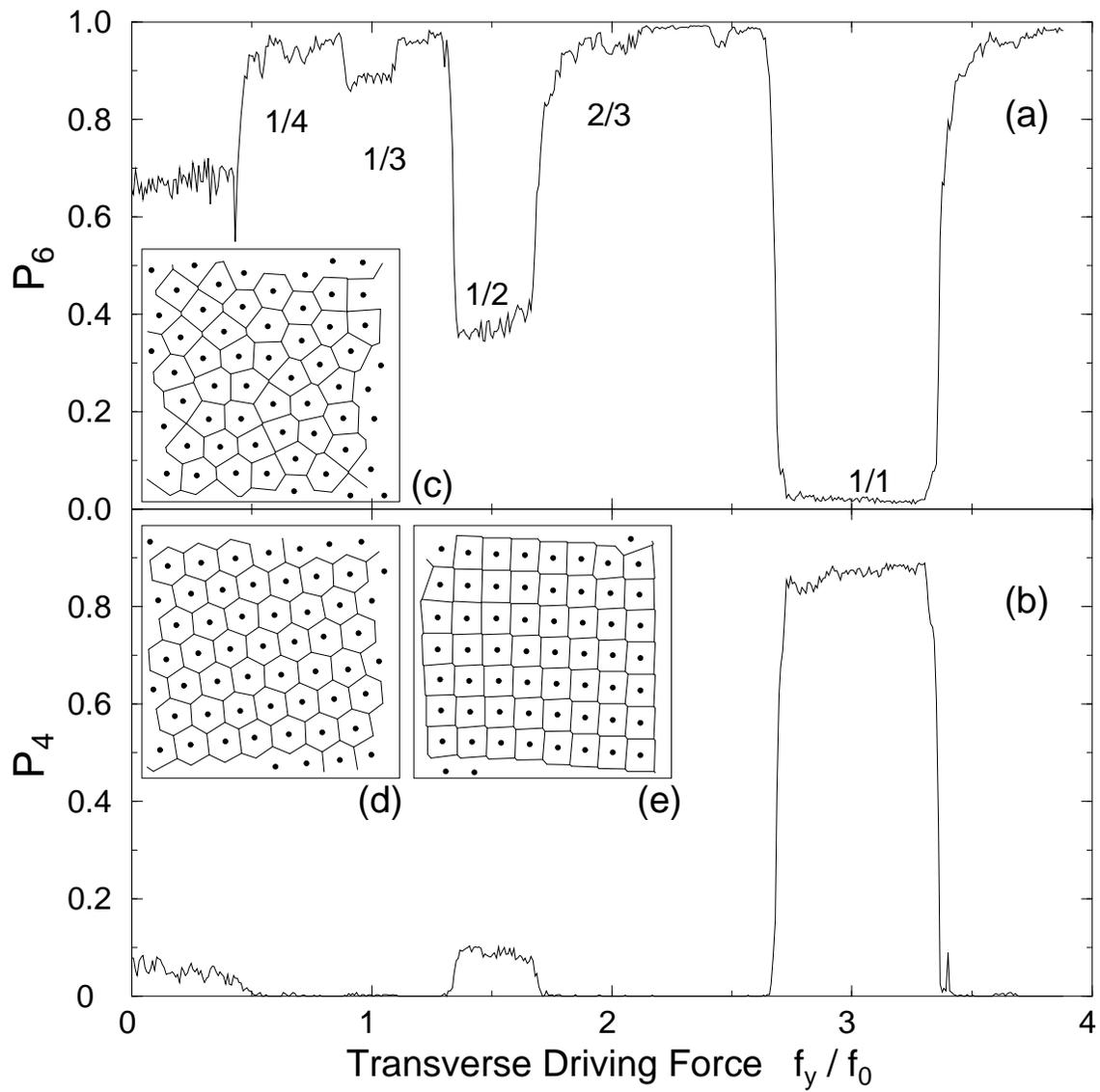
\begin{figure}
\caption{
The fraction of (a) six-fold $P_{6}$ and (b) four-fold $P_{4}$ 
coordinated vortices versus transverse driving force
$f_{y}$, for the same system as in Fig.~1. Large drops in
$P_{6}$ can be seen at $f_{y}^{c}$, 
as well as at the $1/2$ and $1/1$ locking regions.
Smaller dips in $P_{6}$ can be seen at the 
$1/4$, $1/3$, and $2/3$ plateaus regions. At the 
$1/1$ transition $P_{4}$ rises to $ \approx 0.9$ 
indicating a transition to a square VL.
In (c,d,e) both the vortex positions and Voronoi polygons 
for a subset of the VL can be seen for: 
(c) the $1/2$ locking region, 
where a disordered VL is observed;
(d) right before the $1/1$ plateau, 
with a triangular VL;
and (e) at the $1/1$ plateau, 
with a square VL.
}
\label{fig3}
\end{figure}

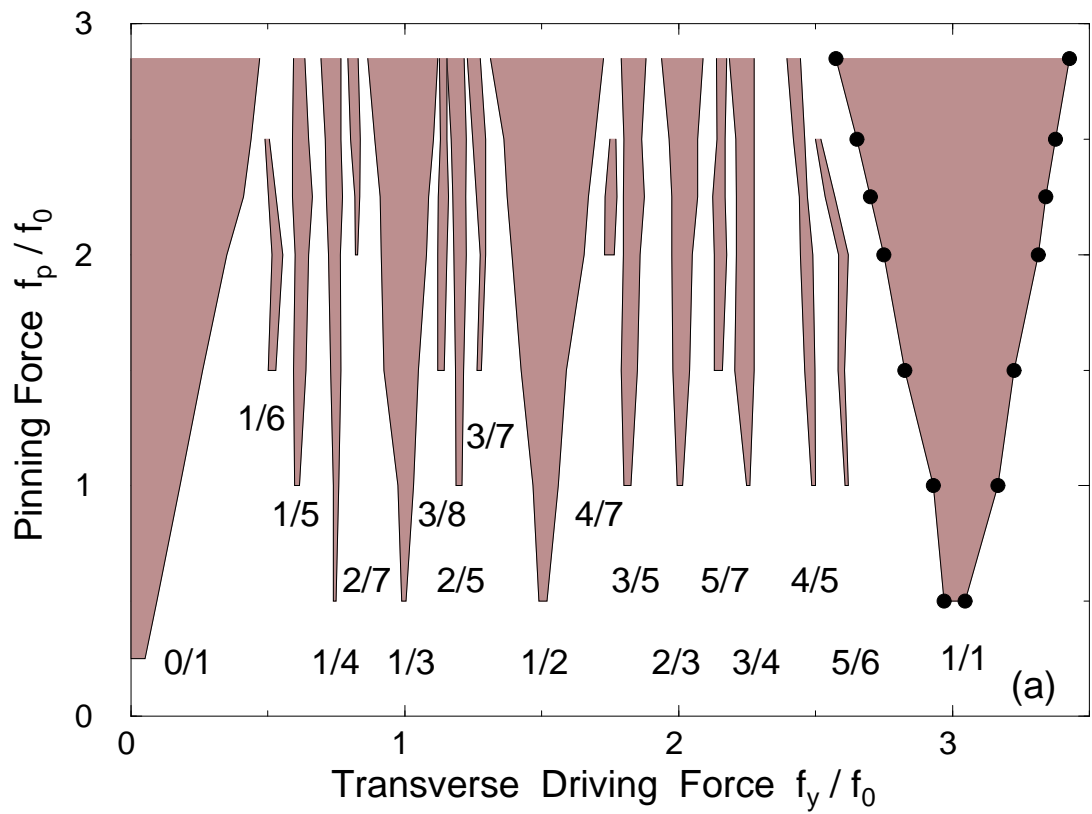
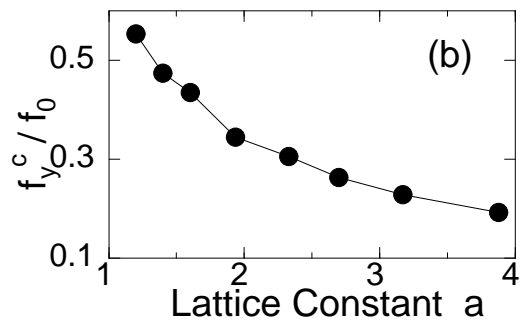
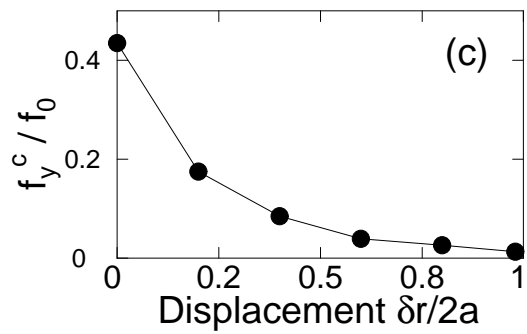
\begin{figure}
\caption{
(a) Phase diagram, for the system in Fig.~1, showing Arnold tongues (shaded); 
i.e., the widths of the locking regions versus $f_{p}$. 
As $f_{p}$ decreases, the tongues (locking regions with
periodic trajectories) shrink. 
In (b,c) the width of the first locking region or $f_{y}^{c}$ is 
shown versus pinning lattice constant $a$ (b) and disorder $\delta r/2a$ (c).
}
\label{fig4}
\end{figure}

\end{document}